# I.T. PROJECT SUCCESS: PRACTICAL FRAMEWORKS BASED ON KEY PROJECT CONTROL VARIABLES

Godfred Yaw Koi-Akrofi[1], Eleanor Afful[2] and Henry Akwetey Matey[3]

[1,2,3]Department of IT Studies, University of Professional Studies, Accra

## ABSTRACT

*The objectives of this study were to research into the interdependencies IT project control variables, and also come out with frameworks to help IT project managers understand how to effectively control these variables to ensure the success of IT projects. The study employed six control variables: Cost, Time (Schedule), Scope, Quality, Risk, and Benefits. A qualitative approach was adopted, where selected IT program and project managers of the Telecom industry in Ghana were interviewed individually and in a group based on a set of questions. The findings, espoused in the frameworks, reiterated the theory of the dependence of one control variable on the other, and the fact that varying one affects the others positively or negatively in relation to IT project success, as is the case for the iron triangle. Again, key activities of the control variables necessary to ensure IT project success were discovered.*

## 1. INTRODUCTION

Information Technology (IT) projects are not too different from all other projects except for the fact that the deliverables/outputs of IT projects are substantially IT artifacts [1]. ICT investment is defined as the acquisition of computer-related equipment and computer software that is used in production for more than one year. ICT has three components: information technology equipment (computers and related hardware); communications equipment; and software. Software includes acquisition of pre-packaged software, customized software, and software developed in-house [2]. IT plays a major role in businesses all over the world today in terms of operational excellence, new products, services, and business models, customer and supplier intimacy, improved decision making, Competitive advantage, and Survival [3]. IT has also become so necessary in today's business due to the following business drivers: Globalization of the Economy, Electronic Commerce and Business, Security and Privacy, Collaboration and Partnership, Knowledge Asset Management, Continuous Improvement and Total Quality Management, and Business Process Redesign [3]. Technology drivers that make IT investments inevitable in today's business include Networks and the Internet, Mobile and Wireless Technologies, Object Technologies, Collaborative Technologies, and Enterprise Applications [3].

Businesses as well as nations have continued to spend a lot of money on IT investments over the years. Table 1 below shows the Worldwide IT Spending Forecast (Billions of U.S. Dollars) by Gartner [4]. Worldwide IT spending was projected to total $3.7 trillion in 2018, an increase of 6.2 percent from 2017, according to the latest forecast by Gartner, Inc. The fastest-growing sectors

are Enterprise Software (8.8 % in 2017, 11.1% in 2018) followed by Devices in 2017 (5.1 %) and IT Services in 2018 (7.4%).

Table 1: Worldwide IT Spending Forecast (Billions of U.S. Dollars)

|  | 2017 Spending | 2017 Growth (%) | 2018 Spending | 2018 Growth (%) | 2019 Spending | 2019 Growth (%) |
|---|---|---|---|---|---|---|
| Data Centre Systems | 181 | 6.3 | 188 | 3.7 | 190 | 1.1 |
| Enterprise Software | 352 | 8.8 | 391 | 11.1 | 424 | 8.4 |
| Devices | 663 | 5.1 | 706 | 6.6 | 715 | 1.3 |
| IT Services | 933 | 4.4 | 1003 | 7.4 | 1048 | 4.6 |
| Communication Services | 1392 | 1.3 | 1452 | 4.3 | 1468 | 1.1 |
| Overall IT | 3521 | 3.8 | 3740 | 6.2 | 3846 | 2.8 |

**Source:** Gartner (April 2018)

IT spending is achieved largely in the form of IT projects in organizations. The success/failure of IT projects over the years has been researched by several researchers tackling it from various angles [5, 6, 7, 8, 9, 10, 11, 12, 13, 14, 15]. The success criteria for a typical IT project are completing the project on time, completing the project on budget, and ensuring that deliverables are according to specifications [16]. Time is of the essence in IT projects, because for example, a company or firm may lose potential and loyal customers to a competitor who may be embarking on the same IT project relating to customers, finishes, and launches early. IT projects are strategic (they are done or initiated to achieve specific organizational/strategic goals in a particular time-frame), and so must be seen as such.

Completing within budget is essential to ensure that the project is not abandoned for lack of funds when cost soars in the course of the project delivery. When deliverables are not according to specifications, project outputs/deliverables do not bring value to the organization; sometimes the deliverables become "white elephants" and are not used or patronized by the intended users. This normally represents a waste of investment in IT.

Despite the high spending in IT investments in organizations worldwide in the form of IT projects, it is recorded in literature that IT projects compared to all other projects record the highest failure rate [17]. This makes the subject of managing successful IT projects key to both researchers and industry players. Several researchers have tackled this subject from different perspectives, all in the bid to put forward models and frameworks to curb this phenomenon. This work intends to tackle this same phenomenon from the perspective of project management control variables.

### 1.1. Objectives of the study

The objectives of the study are:

1. To investigate, among other activities, the key activity required to ensure IT project success, with control variables in perspective.

2. To analyze the level and extent of interdependence of project control variables, and how they impact on IT project success.

The work is also aimed at coming out with practical frameworks to help IT project managers ensure the successful completion of IT projects.

### 1.2. The problem

Despite the high spending in IT investments in organizations worldwide in the form of IT projects, it is recorded in literature that IT projects compared to all other projects record the highest failure rate [11, 17]. This makes the subject of managing successful IT projects key and of great concern to both researchers and industry players [18, 19].

Several researchers have tackled this subject from different perspectives, all in the bid to put forward models and frameworks to curb this phenomenon. This work intended to tackle this same phenomenon from the perspective of project management control variables.

## 2. RELATED AND CONTEXTUAL STUDIES ON IT PROJECT FAILURE/SUCCESS

A lot of factors contribute to the success or failure of IT projects. Studies on Typical sources of IT project failure between and including 1970s and 2000s include: Unclear success criteria, Changing sponsor strategy, Poor project definition, Technology, Concurrency, Poor quality assurance, Poor linkage with sales and marketing, Inappropriate contracting strategy, Unsupportive political environment, Lack of top management support, Inflation, Funding difficulties, and Inadequate manpower [12,13,14, 15].

From a literature-based analysis, Koi-Akrofi [6] found out that four areas are so critical or crucial for the success of IT projects: Skills of project manager/team members, Top Management Involvement, Proper Methodology/Processes, and Good Communication. These were the success factors. For the process dimension, and using the traditional process group or systems development life cycle (SDLC), it was discovered that the success of IT projects depends so much principally on the activities of the initiating stage followed by the planning stage, and then the executing, monitoring and controlling stage, and finally, closing stage.

Other researchers have grouped IT project failures into four categories: Correspondence failure (the information system (IS) fails to meet its design and objectives), Process failure (the IS overruns its budget and/or time constraints), Interaction failure (the users maintain low or non-interaction with the IS), and Expectation failure (the IS does not meet stakeholders' expectations) [7, 8, 9].

The CHAOS study which was published in 1995 by the Standish Group revealed that 53 % of IT projects were completed over budget/schedule which was also referred to as challenged projects; they did not meet all of the project requirements. 31 % of IT projects were canceled before completion, and only 16 % of IT projects were successful. This study gives an idea of the high rate of it project failure.

In a study by Koi-Akrofi, Koi-Akrofi and Quarshie[5], 7 reasons for IT project failures were realized: no project management departments, No quality checks for it projects, IT projects not completed according to schedule, No specific it project methodology followed, Revision of scope very often for a particular project, IT projects not meeting user needs, Wrong estimates.

IT Project failures in organizations in Ghana In analyzing the success criteria of IT projects, one of the key concepts to consider is the concept of control variables. Control variables must be varied appropriately to ensure IT project success. Traditionally, all project methodologies (Project Management Professional-PMP, PRrojects IN Controlled Environments-PRINCE2, and so on) support what is known in project management practice and literature as the triple constraints: project cost, project schedule/time, and project scope. These are the traditional control variables essential for the success of IT projects. The triple constraints put together, form what is also known as the iron triangle [20], and sometimes also called the performance measurement baseline. There are several variations of the triple constraints. Viewing it from the success criteria of IT projects, the constraints become cost, time and quality, but other researchers like Dobson [21], are of the view that the constraints are cost, time and scope, which are inter-related or inter-dependent on each other, and upon varying them, affect quality in the end [22].

Another characteristic of the triple constraints is that varying one affects the other two. When scope is varied, time and cost are affected, and the same is true for the other two. In summary, the triple constraints are scope, time, and cost, which is the same position shared by Dobson [21]. Again, the triple constraints are inter-dependent; that is varying one affects the other two, and finally, varying the three factors, affect quality. This places quality in the middle of the iron triangle.

In PRINCE2's book "Managing Successful Projects with PRINCE2", the concept of control variables has completely been revised to include "Benefits", "Quality" and "Risks" apart from the traditional ones: Cost, Time, and Scope [23]. The proposers of these control variables argue that these six variables are key aspects of project performance that have to be managed always. This is a new twist to the traditional triple constraints concept. The reason for the inclusion of the other three control variables are explained as follows:

"Benefits" is included to ensure that the IT project is not just meeting the success criteria of completing on time, within budget, and built to quality specification, but also ensure that what the project delivers is consistent with achieving the desired returns. The project manager has to have a clear understanding of the purpose of the project as an investment.

Risks are inevitable in project management, but how many risks are we prepared to accept? Thorough analysis must be done to know the allowable limits, mitigation or alleviation plans, and so on, or even if there is nothing that can be done. Projects must work and must be fit for purpose. This is where quality comes in.

In PMP, apart from Benefits which is enshrined in the Initiating process group, the remaining five control variables are treated as knowledge areas which the project team must have in-depth knowledge about. Benefits in PRINCE2 is based more on customer behavior and sophistication. Customers can cause benefits to change based on evolving requirements in the cause of the project delivery, and this must not be overlooked. This behavior of customers makes it inevitable for the fast-evolving method of project agility.

The main objective of this work is to analyze the level and extent of interdependence of these six control variables, and how they impact on IT project success. The work is also aimed at coming out with practical frameworks to help IT project managers ensure the successful completion of IT projects.

Theoretically, the outcome of this work is expected to contribute immensely to the body of knowledge and research work in the success/failure causes of IT projects from the perspective of project control variables.

Practically, the work is expected to help IT project managers know how to do the trade-offs between project cost, project schedule/time, benefits, risks, quality, and project scope while focusing on performance and successful completion of IT projects at the same time.

## 3. METHOD

### 3.1. Research Method

This research adopts purely a qualitative approach. IT program and project managers were interviewed to give their candid views concerning the subject matter. The information gathered from these interviews and discussions formed the basis for the analysis and the construction of the framework.

### 3.2. Research Type

The research format is predominantly Exploratory. Exploratory, because primary research as a method under exploratory was employed. For experience surveys, experienced IT program and project managers were interviewed individually and as a group on the subject matter.

### 3.3. Techniques/tools/approaches/instrumentation/devices

Structured questions were posed to the IT project and program managers individually and in a group. A discussion format was employed for both the individual interview and the group interview. Notes were taken in the course of the discussions, as well as recordings on mobile phones.

### 3.4. Data collection methods

The study was a case study with the Telecommunications industry in Ghana as the object of study, and primary data was solely used for the analysis. No secondary data was used.

### 3.5. Population and sampling procedure

The study employed non-probability sampling methods of quota and convenience. Quota in the sense that representative individuals were chosen from a subgroup as against random sampling of members of the group. In this particular instance, the study focused on the IT program/project managers as representatives from the project management offices (PMOs), which are units/subgroups under the Technology departments of the companies in the telecom industry in Ghana. The reason for the researcher was that the subject of the research is a specialized area that may be best understood by practitioners, and not anybody working in the PMO unit or Technology department. It is also convenient because the companies in the industry were easily accessible to the researcher and also easier to reach in terms of proximity.

In summary, the population employed was IT project and program managers of the Telcos in Ghana, and the sample size was two IT project managers and 1 program manager for each of the Telcos, making up to 12 people.

## 4. RESULTS ANALYSIS

Three major questions were asked the IT program and project managers. These questions are:

1. Practically, what is the key activity that is done concerning each control variable in the course of IT project delivery to ensure project success?

2. Practically, what are the level and extent of interdependence of the six project control variables on each other? And

3. Practically, what are the general precautions to take concerning each project control variable to ensure IT project success?

The premise for the interview with the program/project managers was agreed as follows:

1. That there are six project control variables namely cost, time, scope, quality, benefits, and risks.

2. That the success criteria for IT project is completing on time, completing within budget, and delivering according to specifications

### 4.1. Project Time

The first question was asked with "time" as the first control variable considered. All the respondents emphasized the fact that "time" is one of the main attributes of an IT project. A project has a beginning and an end; a project, especially in an organizational setting, is not meant to last for a lifetime, because it is normally delivered to address a pressing need which is strategic and time-bound. It was agreed that for time to be adhered to in IT project delivery, the key activity that must be looked at constantly in the course of the project delivery is for the project manager to ensure "scheduled/random/regular monitoring and reporting including progress reporting" (See Figure 1). Scheduled/random/regular monitoring will put the team on their toes to deliver on time as is agreed from the beginning. Again, regular reporting including progress reports will inform the project manager as to the pace of delivery, the reasons for that pace, and act accordingly to ensure that the project is completed within the stipulated time. Some of the actions the project managers can do to redeem project time are quick and early escalation where necessary, initiating change requests, resolving project issues such as conflicts among stakeholders, and so on.

On the dependencies (see Figure 2) of time on other control variables, this is what the respondents had to say: for project risk, they maintained that planned risks chosen from a bucket of anticipated risks are manageable, and do not normally affect time, because they are factored or catered for in the determination of the time for the completion of the project right from the beginning. Unplanned risks, however, can throw overboard the project time since they are not anticipated.

An example of unplanned risks is "force majeure". It was also discovered that when project-related issues such as conflicts, technical problems, and so on, which normally do occur in the course of the project delivery, are not managed well, project time can be affected. Risks are futuristic, whiles issues are current or immediate. You look forward to risk, but issues (forms of risks being managed now) have to be resolved to ensure the project comes back on track as quickly as possible. A typical example is when the project manager is bent on following governance structures strictly against other stakeholders who may not want to obey them.

Cost is a major control variable that can also affect project time. Once the project runs out of budget due to overspending or cut of project funding inflow, the project can come to a halt or project time can be prolonged unduly in the absence of funds to immediately cover costs.

Scope creep is unplanned scope changes or revisions and can affect project time if allowed to go on. When scope changes are planned, revisions go through change requests, and so are agreed by all stakeholders, and if project time is affected, it does not become an issue because it is also agreed unanimously by all stakeholders.

On benefits, the respondents made it clear that, in the bid to convince stakeholders of the need to do the project at all cost, benefits are sometimes exaggerated. If benefits are exaggerated and are later on found not to be so, it may slow down the IT project entirely. This is because stakeholders may not be too interested in the output as it used to be from the beginning. Apathy and disaffection for the project can cause it to delay.

On quality, the respondents said that when quality is compromised, change requests and reworks to remedy the situation may result in IT project delay.

When a project is not completed on time, it is considered as not being successful. This is because a project is time-bound; after that stipulated time, the project may not be useful to both customers and the organization. For IT projects, time is critical, because the deliverable or output may not serve the intended use after a stipulated time.

### 4.2. Project Scope

The respondents summarized one key activity that has to be done at the very early stages of the IT project to ensure that scope is not revised unnecessarily; and that is, "ensuring that

International Journal of Software Engineering & Applications (IJSEA), Vol.10, No.5, September 2019

requirements gathered are adequate and complete based on the objectives and the benefits of the project" (See Figure 1). Once requirements are complete, there is a very slim probability that scope will be revised in the course of the project. Inadequate requirement gathering always poses the chance for the team to revise scope when they are confronted with new requirements. Scope creep should be completely discouraged, but if there is a genuine need for scope revision, it would have to be agreed upon by all stakeholders, and it must go through a change request for approval before it is effected, so that all the other dependencies such as cost, and so on can be collectively and properly addressed.

Upward revision of scope means additional work and resources, which have a bearing on the cost. Additional work suggests an increase in project time if the same resources are used unless the team works extra to recover time. If resources are beefed up because of the additional work, it will increase costs in the end.

Additional work may introduce new risks and issues into the process, and so new risks can also occur due to scope revision. Additional work may also compromise quality, especially, in the situation where stakeholders are not prepared to vary the key dependencies (cost, time) to correspond to the changes in scope (see Figure 3).

From the discussion so far, it is obvious that revision of scope can affect the success of IT projects in terms of completing within time, completing within budget, and deliverables being according to specifications (quality).

## 4.3. project Cost

The respondents unanimously mentioned that "initial cost estimates" is key for cost to stay within budget (See Figure 1). If initial cost estimates are not done well, actual costs later in the course of the project may throw the budget into disarray. In estimating costs, indicators such as inflation, exchange rate, interest rate, and so on must be taken into consideration. Normally, current indicators are not used for estimation. Extrapolation of indicators must be made to ensure that at the time of procuring items for the project delivery, estimates done initially will almost tally with actual prices of items in the market.

Revision of scope affects project cost. Upward revision of scope, whether by scope creep or through a change request introduces additional work, which may require additional resources such as human resources, logistics, and so on, which are all cost items. When project time is prolonged, cost increases because additional human resources may be needed to expedite the completion of the project to meet organizational goals. Quality issues may arise from deliverables not conforming to specifications. When this happens, re-works are inevitable, and re-works always obviously increases the cost. From the discussion, it is obvious that when project cost goes up, the success of IT projects can be affected in terms of completing projects within budget, completing within time, and completing projects with deliverables according to specifications (See Figure 4). Most of the time, anticipated or planned risks should not affect project cost because it is always catered for, but the unplanned risk may affect project cost negatively.

**4.4. Project Quality**

Quality in IT projects is very essential because intended users must be happy with the outcome of the project in terms of the functional and non-functional requirements the developer outlined and put together with the help of the intended users at the early stages of the project. Quality means that the output of the project must serve the very purpose for which it was made. The respondents unanimously agreed that for IT quality to be ensured, there must be "efficient and effective verification and validation processes during the development process" (See Figure 1). These processes include inspection (especially requirement inspection), various forms of testing, and effective monitoring and controlling. Through monitoring and controlling, any deviations from the original specifications can always be corrected either with or without change requests.

When the project is delayed, there is a high tendency to compromise on quality to finish on time. When project spending soars and there is the need to check costs or minimize cost, it tends to affect quality negatively, regardless of how the project manager tries to avoid it. This is because there may not be enough financial resources to do what is expected or required to do. Upward revision of scope affects quality negatively, especially when it does not go through the right procedure involving change request. The project manager in order not to let stakeholders be aware of scope creep, may conceal it, and eventually compromise on quality since cost and time are also affected in the process. Risk may also affect quality negatively, especially when the risk is not predicted and it happens anyway. There may be so much pressure on the team to deliver before a particular day, and this may cause them to put all caution to the winds (in terms of quality) and deliver regardless. Exaggerated benefits can cause project quality to be compromised as the team may want to do other things like cutting down cost for some planned project activities to ensure that they can do some other activities which they think will enhance project benefits (See Figure 5).

**4.5. Project Risk**

On project risk, the unanimous decision from the respondents was that, when risks occur, they can always be managed well, if "risks are anticipated from experience knowledge and organizational process assets" (See Figure 1). A list of risks from a bucket of risks are normally chosen by the project manager for a particular project, and practically, it is normally based on prior knowledge of risks that are associated with past similar projects. The best way to deal with project risk is to avoid it entirely. If it becomes difficult to avoid its occurrence, then there must be a detailed program to manage it so it will not have any negative effect on the success of the project. Project risk is contextual; that is to say that, what is seen as a risk for a particular jurisdiction may not be risk elsewhere. In general, project risk should be tied to project objectives. In that sense, project risk can be defined as an uncertain event or condition that, if it occurs, affects at least one project objective (https://pm4id.org/chapter/11-1-defining-risk/). From this definition, project delays, increase in cost, scope creep, and so on can all be referred to as project risk, because they can affect the IT project success criteria, which are basically, the main objectives of every IT project (See Figure 6).

**4.6. Project Benefits**

Project benefits are situated in justification for the IT project. This is done at an early stage of the project management life cycle. The main aim is to determine the investment benefits of the project to decide whether it is worth doing or not. The benefits must outweigh the cost to pass for a viable project. For tangible benefits, Cost-Benefit analysis is normally used for this exercise, and some of the evaluation methods employed are the payback period, net present value, internal rate of return, profitability index, and so on. Practically, the respondents argued that intangible benefits are taking over from the tangible benefits as the consideration for the viability of, especially, IT projects (See Figure 1). In his work on "Justification for IT Investments: Evaluation Methods, Frameworks, and Models", this is what Koi-Akrofi [6] had to say about intangible benefits:

> "Intangible benefits are difficult to identify from the beginning and once identified, difficult to measure/quantify, and for that matter difficult to justify/evaluate. Tangible benefits are not enough to justify for IT investments as they mostly point to corporate benefits which are short term, and that intangible benefits justification must be included to make the justification process complete no matter how difficult it is. Again, intangible benefits are usually strategic in nature, requiring some time for realization, and must be seen as such; there should be no rush".

In determining intangible benefits, the customer must be the focus. Sometimes, all the evaluation methods may point to the fact that the project is not viable in terms of the finances to the company or firm, but once it can be proved that customers will benefit from it, the project must be done. To buttress this point, Koi-Akrofi [6] in his work "Justification for IT Investments: Evaluation Methods, Frameworks, and Models", had this to say:

> "Sometimes the firm should allow certain benefits, even though not a direct benefit, during the decision-making process to overcome the financial considerations (direct benefit). An example of such is customer satisfaction. This may not be easily measured, but can benefit the firm in the long term, and may translate into financial benefits in the end".

Project benefits may change in the course of the project delivery. This normally happens due to the sophisticated nature of say, customers, when it comes to technology products. Customers are the ultimate users or beneficiaries of the IT project, and so their preferences must be monitored and factored in the project delivery process, to ensure full patronage of the product for a long time, to guarantee financial benefits for the firm in the long term. IT project delay can erode benefits if benefits are short term. For instance, if customers are to benefit from the output or deliverable in a stipulated time, delay can cause the output to be irrelevant. Overspending or cut in the flow of funds can cause delays if funds are not immediately available, and can cause the same problem as above. A decrease or reduction in spending may not affect project benefits. Planned upward scope can cause the cost to soar, as well as delay in time, and may hurt short time benefits, but for the fact that it is planned, all these problems are supposed to be mitigated. Planned downward scope revision may or may not affect benefits. It may because it can reduce expectations of stakeholders or even intended users/customers who had prior knowledge of the original scope. Expectations can translate into benefits where applicable. Scope creep upward will affect benefits negatively. Scope creep downward also may or may not affect benefits depending on the scenario at hand. Poor quality of work will affect benefits as patronage in deliverables may go down to affect revenues, and so on. For anticipated or planned risk, most of the time should not affect benefits because it is always catered for. Unplanned risk, however, may affect project benefits severely (See Figure 7).

International Journal of Software Engineering & Applications (IJSEA), Vol.10, No.5, September 2019

The summary of the discussions are captured in the frameworks below:

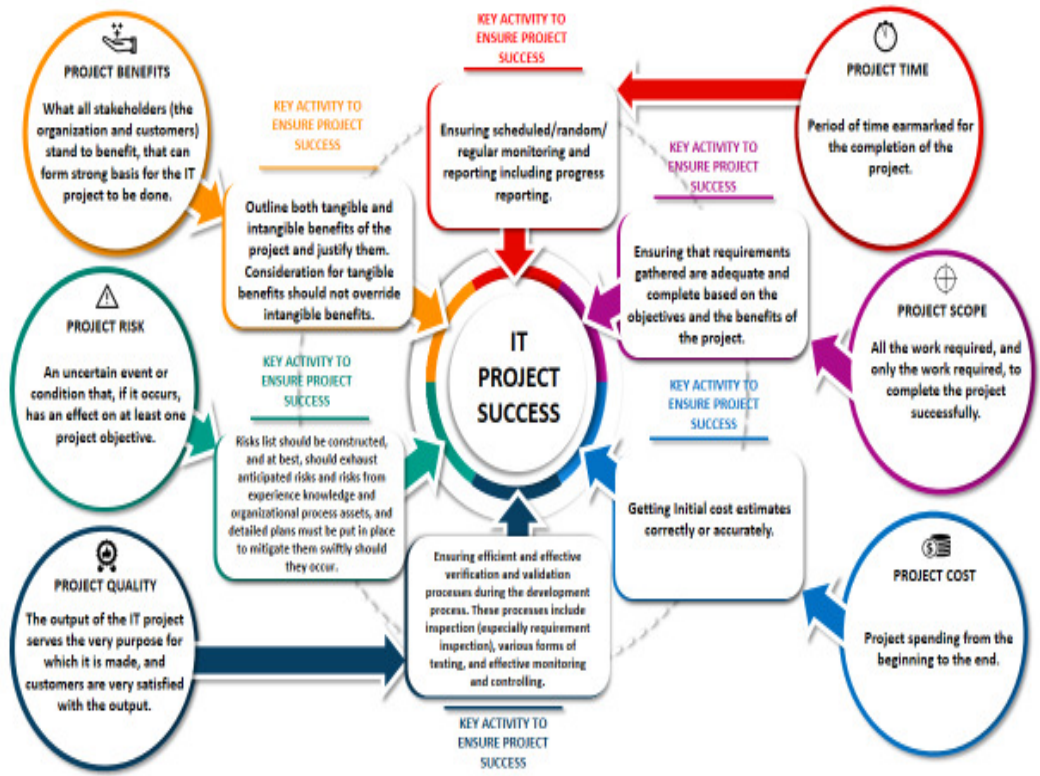

**Figure 1:** IT Project Success and Control Variables Framework (main)

From Fig. 1, the control variables are described (in the circles), and the key activities concerning each control variable to ensure IT project success (in the rectangles), are also indicated.

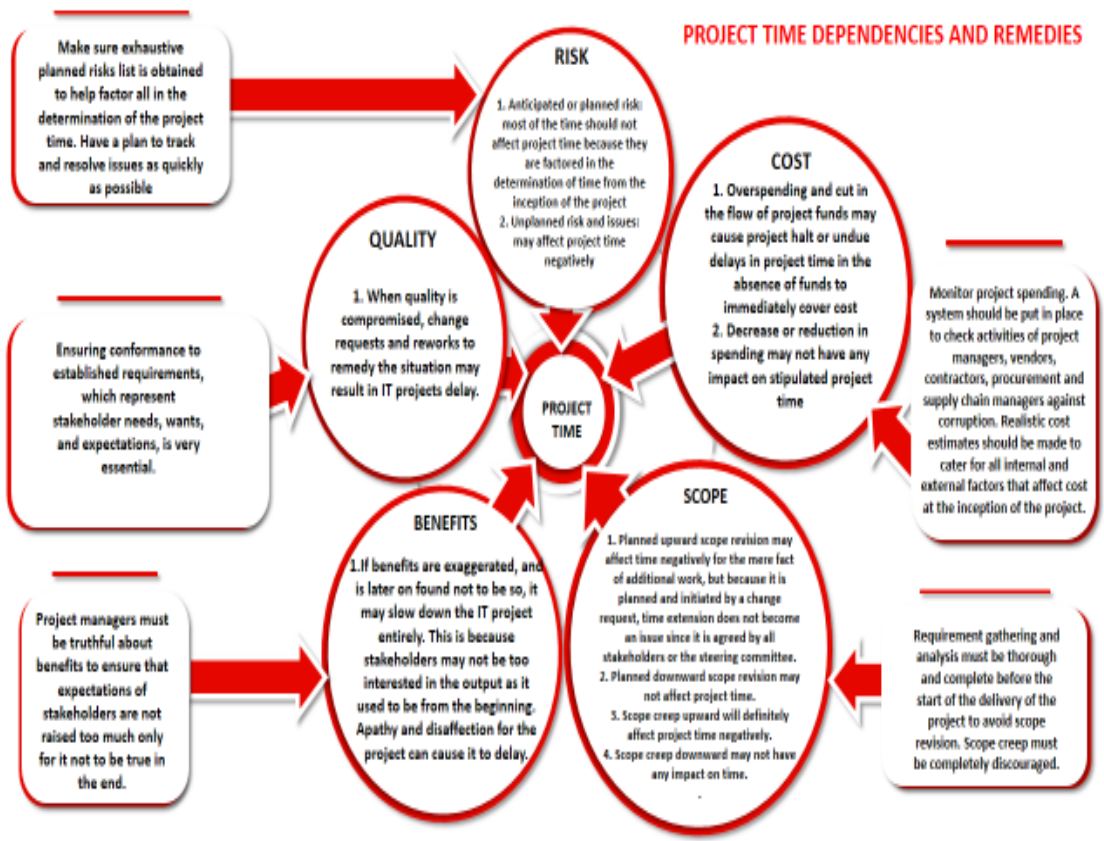

**Figure 2:** IT Project Success and Control Variables Framework (Time dependencies)

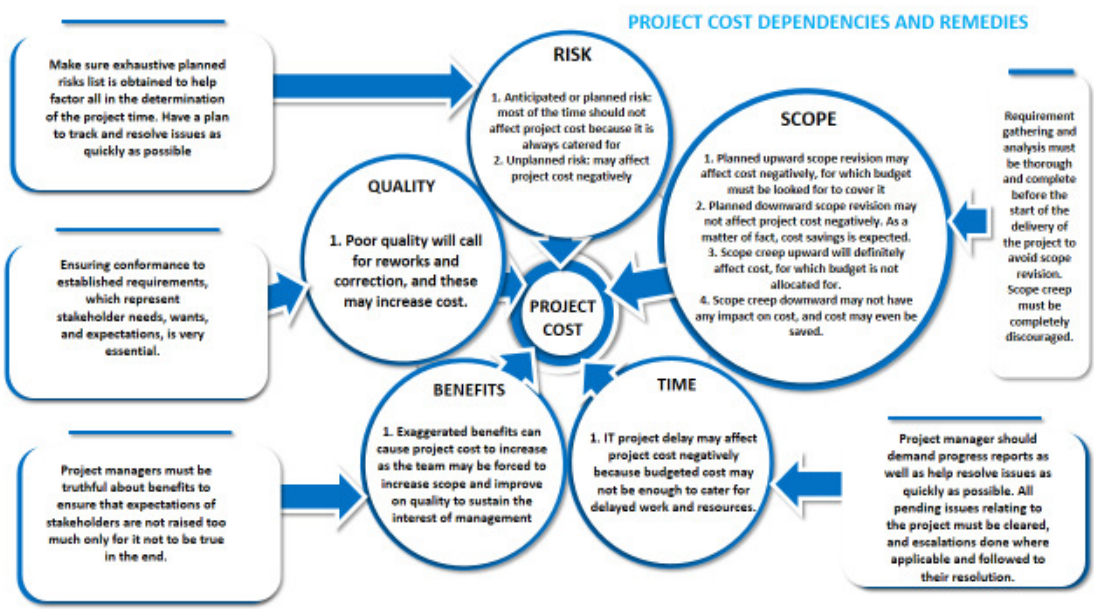

**Figure 4:** IT Project Success and Control Variables Framework (Cost dependencies)

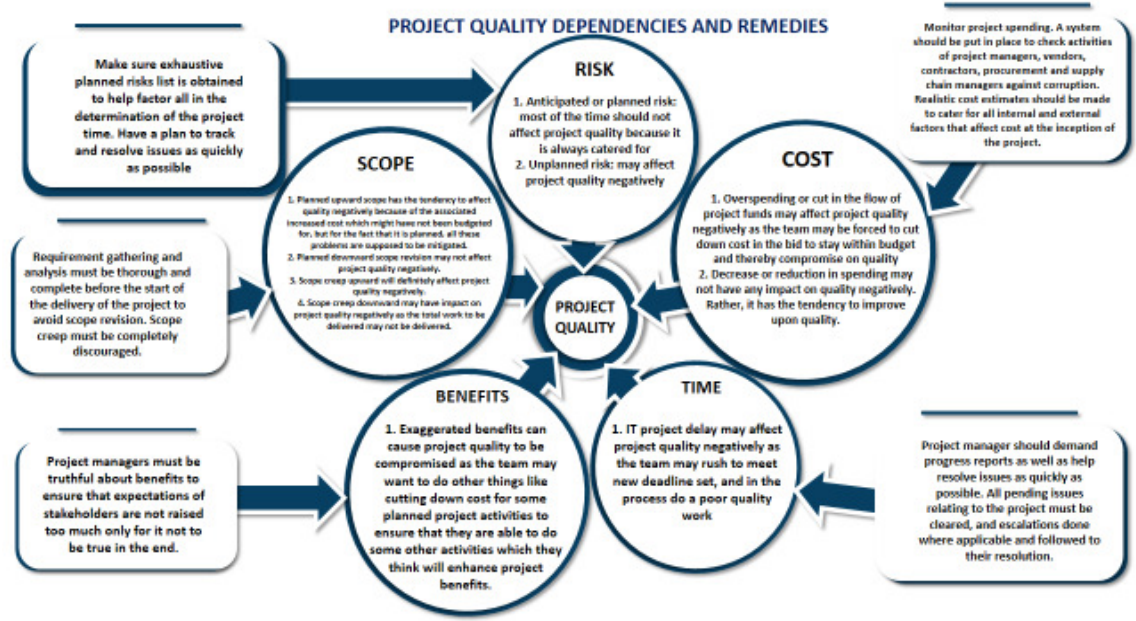

**Figure 5:** IT Project Success and Control Variables Framework (Quality dependencies)

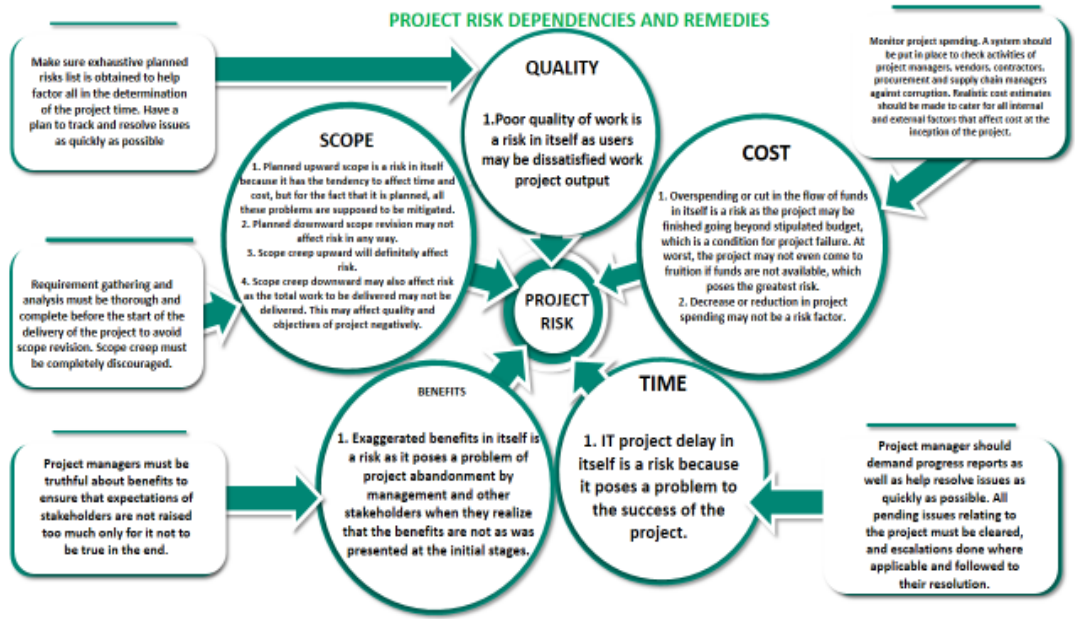

**Figure 6:** IT Project Success and Control Variables Framework (Risk dependencies)

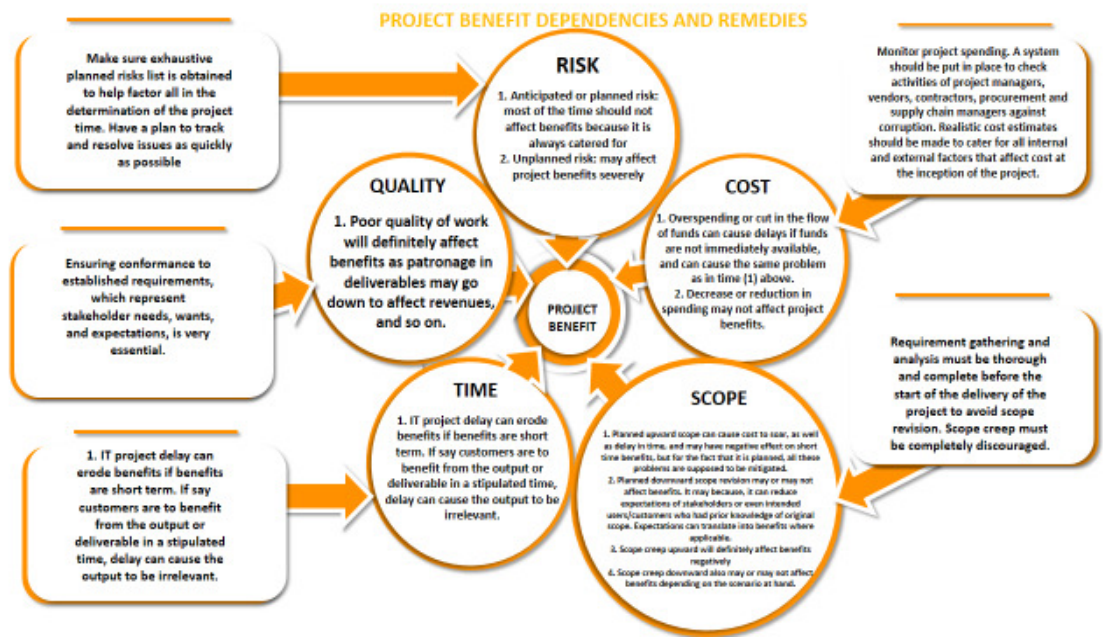

Figure 7: IT Project Success and Control Variables Framework (Benefit dependencies)

## CONCLUSION

Several factors can contribute to the failure of IT projects. This research work focused on how control variables namely cost, risk, time, benefits, quality, and scope impact on project success. It was established from the output of the research that these control variables are interdependent. Altering any of the six control variables will impact positively or negatively on the other five and vice versa.

Framework to describe this dependency relationship and its effect on project success have been developed. The frameworks are a single reference point that can be used by IT project managers to know exactly what to focus on to achieve project success based on the six control variables.